\documentclass[twocolumn]{aastex701}
\usepackage{amsmath}     
\usepackage{wasysym}           
\usepackage{graphicx}
\usepackage{amssymb}
\usepackage{epstopdf}
\usepackage{mathrsfs}
\usepackage{anyfontsize}
\usepackage{natbib}
\usepackage{makecell}
\usepackage{enumitem}
\usepackage{color}
\usepackage{lipsum}
\usepackage{diagbox}

\definecolor{purple}{HTML}{AD26FB}

\shorttitle{Wind-generated Fireworks}
\shortauthors{Coughlin, Salvesen, \& Pasham}
\begin{document}
\title{A Wind-Driven Origin for the Firework Morphology of the Supernova Remnant Pa 30}

\author[orcid=0000-0003-3765-6401]{Eric R.~Coughlin}
\affiliation{Department of Physics, Syracuse University, Syracuse, NY 13210, USA}
\email[show]{ecoughli@syr.edu}
\correspondingauthor{Eric Coughlin}

\author[0000-0002-9535-4914]{Greg Salvesen}
\affil{Center for Theoretical Astrophysics, Los Alamos National Laboratory, Los Alamos, NM 87545, USA}
\email{}

\author[0000-0003-1386-7861]{Dheeraj R.~Pasham}
\affiliation{Eureka Scientific, 2452 Delmer Street Suite 100, Oakland, CA 94602-3017, USA}
\affiliation{Department of Physics, The George Washington University, Washington, DC 20052, USA}
\email{}

\begin{abstract}
Pa 30---the likely remnant of the Galactic type Iax supernova of 1181 AD---displays an unusual, firework-like morphology, consisting of radial filaments extending from a common center, where a white dwarf (WD) currently drives a very fast wind (speed $\gtrsim 10^{4}$ km s$^{-1}$). We propose the filaments arose from the Rayleigh-Taylor-unstable nature of the interface between the circumstellar medium (CSM) and the shocked wind launched by the natal WD; the filaments then elongated intact due to the Kelvin-Helmholtz-stable nature of the large initial density contrast between the wind and CSM, supplemented by the slowly declining wind density profile (relative to homologously expanding ejecta). To support this interpretation, we present two-dimensional hydrodynamical simulations and derive the filament properties, including their speed, density, and temperature, all of which are consistent with observations. We suggest the filaments elongate until the wind and CSM densities become comparable at the contact discontinuity, which occurs within 1--10 years, and then truncate because the RTI halts. The subsequent KHI growth timescale across the current width of the filaments is longer than the age of Pa 30, so they remain intact. The filament-less central region in Pa 30 is therefore more likely a consequence of the finite timescale over which the RTI operates, rather than a wind termination shock. In general, firework-like filaments may form in other systems, provided there is a sufficiently large density contrast between the ejecta and its surroundings.
\end{abstract}


\section{Introduction}
The supernova remnant (SNR) Pa 30 \citep{gvaramadze19, oskinova20, ritter21, schaefer23} is tentatively linked to the historical Galatic supernova of 1181 AD \citep{hsi57}, placing its age at ${\sim}850$ yr. This SNR is remarkable in (at least) three respects: 1) the morphology of nearly radial ``spikes'' extend from a seemingly common center; 2) at this center is a white dwarf (WD) remnant, which is unexpected for most evolutionary pathways of type-Ia supernovae (e.g., \citealt{whelan73, iben84, nomoto84, howell01, scannapieco05, shen10, kashi11, hillebrandt13, maoz14, polin19, ruiter25, shen25}; but see \citealt{foley14, mccully22}), even if they are a relatively common byproduct of low-mass WD mergers (e.g., \citealt{badenes12, shen12, temmink20, kilic24}); and 3) the WD launches a fast wind \citep[$\gtrsim 10^4$ km s$^{-1}$;][]{gvaramadze19, garnavich20}, plausibly driven by magnetohydrodynamic effects (e.g., \citealt{gvaramadze19, kashiyama19, zhong24}). These features---especially the radial filaments---are atypical of young, type-Ia SNRs. 

A natural supposition is that these features of Pa 30 are related, and \citet{fesen23} argued that the wind from the WD remnant accelerated the supernova ejecta, which then evolved into filamentary structures through the Rayleigh-Taylor instability (RTI) and wind ablation. Lending support to this interpretation, other (non-SNR) systems with winds display overdense and somewhat elongated structures (e.g., \citealt{vaytet07, shara12}); however, those structures display mass loss, as expected if they are forming in the wake of an ablative wind, yet no such mass loss has been observed from Pa 30 \citep{fesen23, cunningham24}. \citet{fesen23} also required the interaction between the wind and the supernova ejecta to be subsonic because supersonic interactions typically lead to short (rather than long, as in Pa 30) tail-like structures \citep{pittard05}; however, Pa 30's high-speed wind ($\gtrsim10^{4}$ km s$^{-1}$) should be highly supersonic.

Challenging the \citet{fesen23} wind hypothesis, \citet{duffell24} asserted that the interface between a wind and its surrounding medium should give rise to the Kelvin-Helmholtz instability (KHI), breaking up the RTI-induced filaments and leading to a turbulent morphology.\footnote{\citet{duffell24} cited the Crab SNR, which hosts a central pulsar wind nebula, as an example of such complex morphology; however, a pulsar wind versus a WD wind could be qualitatively distinct, as could the surroundings into which they advance.} This motivated \citet{duffell24} to hypothesize that the Pa 30 filaments arose from anomalously strong radiative cooling of the shocked supernova ejecta, leading to compression and KHI stabilization; hydrodynamical simulations of artificially cooled, homologously expanding material supported this interpretation \citep{duffell24, pikus25}.

Both hypotheses above presume the original supernova ejecta---acting either by itself \citep{duffell24} or in tandem with the WD wind \citep{fesen23}\footnote{See also \citet{ko24}, whose model requires a late (c.~1990) turn-on of the WD wind and its subsequent interaction with the supernova ejecta to explain the Pa 30 observations.}---led to the formation of the filaments and make up their composition. However, type Iax supernovae are weak (compared to type Ia), and radio non-detections constrain the Pa 30 explosion energy to be $\lesssim {\rm few} \times 10^{47}$ erg \citep{shao25}, which is 3--4 orders of magnitude below a typical type Ia. \citet{shao25} also note that the kinetic energy from the WD wind is ${\sim}10^{48}$ erg if it remained active for ${\sim}100$ years, and this energy could be even larger if the wind was more powerful just after the merger (as simulations suggest, e.g., \citealt{shen17}; see also \citealt{fesen23}). This energy inventory implies that the wind itself could rival or exceed the dynamical importance of the explosion in shaping the nebula, or perhaps---as \citet{shao25} suggest---Pa 30 is actually a wind-blown nebula, not an SNR. 

Instead, we propose that the WD wind (not the original supernova ejecta) is responsible for the Pa 30 firework morphology, as follows from RTI and KHI considerations. We expect the wind density was initially much larger than the circumstellar density (see Section \ref{sec:parameters}). In this case, the RTI growth rate is relatively insensitive to the density contrast between the wind and its surroundings (see Section \ref{sec:stability}). Contrarily, the KHI growth rate scales as the square root of the ratio of the less-dense to the more-dense fluid. Thus, the KHI should be inhibited when the density contrast is large, despite the RTI-induced velocity shear. The shear itself is also minimized during this phase, because the forward shock moves out only marginally faster than the contact discontinuity, which further reduces the KHI growth rate. Moreover, the wind density declines relatively gradually (as $\propto R^{-2}$ versus $R^{-3}$ for a homologous explosion) with distance $R$ to the leading edge of the expanding wind, which further serves to maintain a high density contrast and stabilize the RTI-induced plumes against the KHI. In summary, the interaction of a dense wind with a diffuse surrounding medium should be unstable to the RTI but relatively stable to the KHI.

\begin{figure*}
    \includegraphics[width=0.995\textwidth]{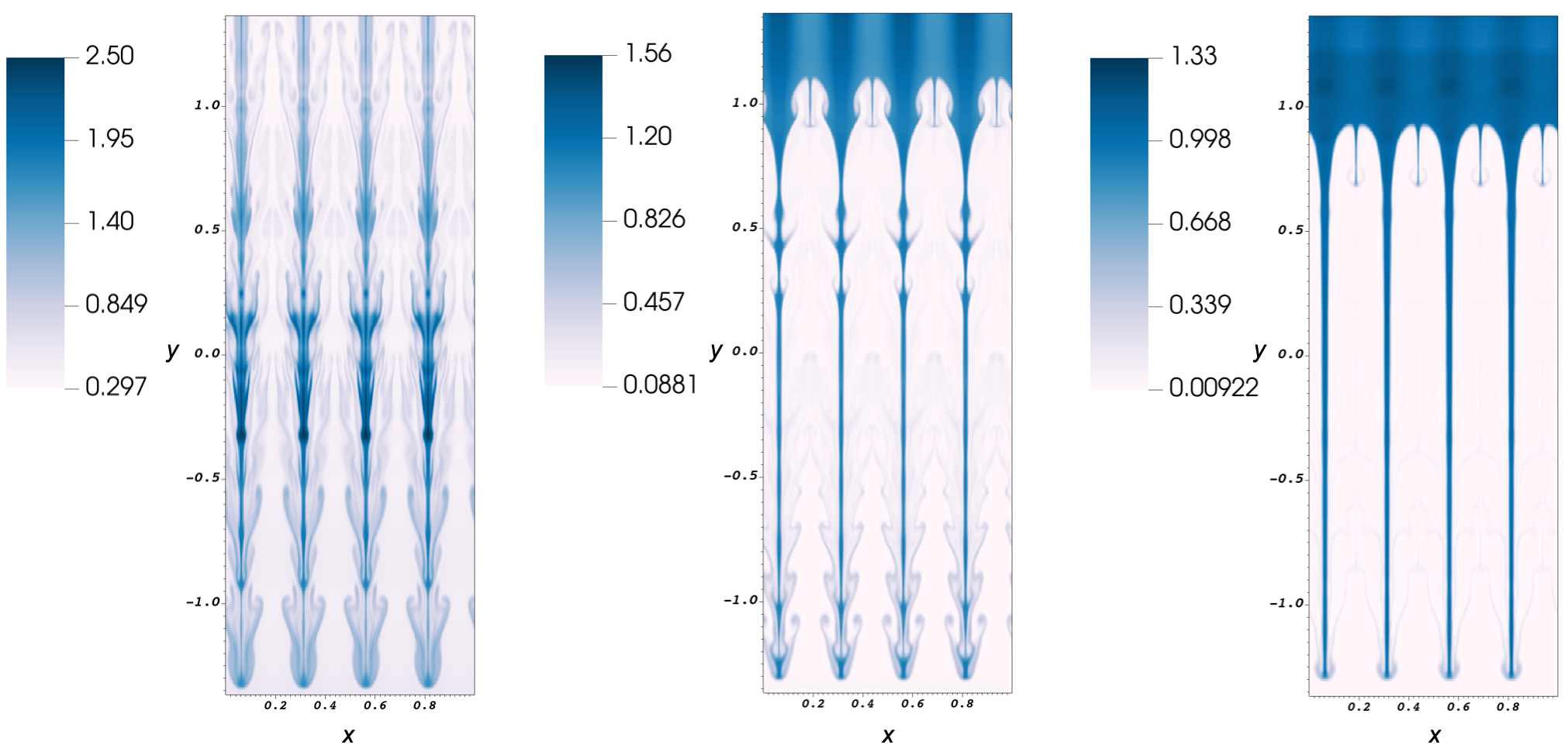}
    \caption{Development of the RTI between an upper fluid with a density of 1 (in code units) and a lower fluid with a density of $0.5$ (left), 0.1 (middle), and 0.01 (right). The fluids are initially separated by a contact discontinuity at $y = 0.5$ and placed in a uniform gravitational field (magnitude $g = 1$; direction $-\hat{y}$), and an initial velocity perturbation of $-0.03 \sin \left(8\pi x\right)$ is imposed. The times in each panel correspond to when the downward-propagating plumes reach the bottom of the domain at $y \sim -1.5$.} 
    \label{fig:rti}
\end{figure*}

As a simple demonstration of this notion, Figure \ref{fig:rti} shows three {\sc flash} \citep{fryxell00} simulations of the RTI development between two fluids, initially separated by a contact discontinuity at $y = 0.5$ and in hydrostatic equilibrium with a gravitational field $\mathbf{g} = - \hat{y}$. Initially, the upper fluid has a density of 1; the left, middle, and right simulations have a lower-fluid density of $0.5$, $0.1$, and $0.01$, respectively. In each case we imposed a velocity perturbation of $\mathbf{v}=-0.03\sin\left(8\pi x\right)\hat{y}$ to seed the instability. The left-most simulation gives rise to KHI-induced rolls alongside the RTI plumes; these tendrils tend to shear apart, mixing the high- and low-density fluids. However, the simulations with larger density contrasts lead to relatively unperturbed ``spikes'' of preferentially downward-propagating fluid. 

We thus propose that Pa 30's unusual firework morphology was produced by the RTI acting primarily on the interface between the (shocked) WD wind and the (shocked) surrounding medium, and that the dense spikes consist of shocked wind material. The resistance to the KHI---and the preservation of coherent RTI spikes---is provided by the high density contrast and low velocity shear between the wind and its surroundings, obviating the need for excessive line cooling invoked in \citet{duffell24}. The large density contrast and the slowly declining radial density profile of a wind also keep the Rayleigh-Taylor unstable contact discontinuity in close proximity to the forward shock for considerably longer than in the case of a homologous explosion, allowing the RTI-driven spikes to reach large radii at early times. We develop this model below by estimating the growth rate of the RTI in the scenario of a dense wind impacting a constant-density medium; then we delineate the properties of the RTI-induced spikes, showing that they are consistent with observations.

\section{Wind-driven Rayleigh-Taylor instability}
\label{sec:wind}
\noindent
The WD wind interacts with a gaseous reservoir, which we interchangeably refer to as the ambient medium or circumstellar medium (CSM), generated by prior phases of stellar evolution and subsequently contaminated by the type Iax explosion ejecta; the interstellar medium (ISM) at $\gtrsim$ pc scales does not enter prominently into our analysis. The term wind refers to the unshocked material being launched at a $\sim$constant speed from the WD remnant. The reverse shock (RS) and contact discontinuity (CD) bound the shocked wind; the forward shock (FS) and CD bound the shocked ambient medium (or shocked CSM).

\subsection{Model parameters}
\label{sec:parameters}
\noindent
Observations of Pa 30 place the WD wind speed at ${\sim}10^{4}$ km s$^{-1}$ and its mass loss rate at ${\simeq}10^{-6}{-}10^{-7} M_{\odot}$ yr$^{-1}$ \citep{lykou23}. These WD wind properties lead to comparable initial densities across the CD that would not be conducive to RTI-induced filament formation (see below and Section \ref{sec:stability}). However, the natal WD merger remnant likely had a lower wind speed $V_{\rm w} \lesssim 3\times 10^{3}$ km s$^{-1}$, and a higher mass-loss rate $\dot{M} \simeq 10^{-5} M_{\odot}$ yr$^{-1}$ \citep{shen17}; 
we adopt these as fiducial initial values that applied near the supernova epoch. Notably, the initial mass-loss rate may have been even larger owing to the likely ONe-CO progenitor \citep{oskinova20}, the very low-energy nature of the explosion \citep{shao25} plus substantial fallback of the initial ejecta, and the corresponding possibility of an initially hyper-accreting disc launching the wind (instead of the delayed radioactive decays as described in \citealt{shen17}); we revisit this possibility at the end of Section \ref{sec:conclusions}. Scaling the radius $R$ to the WD surface at $0.01 R_{\odot}$, the wind density profile is
{\small 
\begin{align}
&\rho_{\rm wind}(R) = \frac{\dot{M}}{4\pi R^2 V_{\rm w}} \label{rhow} \\
&\simeq 3{\times}10^{-7} \frac{\textrm{g}}{\textrm{cm}^{3}}\left(\frac{\dot{M}}{10^{-5} \frac{M_{\odot}}{\textrm{yr}}}\right) \left(\frac{R}{0.01 R_{\odot}}\right)^{-2} \left(\frac{V_{\rm w}}{3{\times}10^{3} \frac{\textrm{km}}{\textrm{s}}}\right)^{-1}. \nonumber
\end{align}
}

Presumably, this WD wind impacts an ambient medium composed of supernova ejecta and material from previous evolutionary phases (e.g., common envelope, asymptotic giant branch). Thus, we let the ambient medium with which the wind interacts start at a radius of $R_{\rm i} \simeq 10^{15}$ cm \citep[i.e., a factor of ${\sim}10{-}100$ larger than the radius of a typical red supergiant;][]{levesque17} with a number density\footnote{The number density at distances from the WD $\gtrsim 1$ pc, which is roughly the outer-tip location of the Pa 30 filaments \citep{fesen23, cunningham24}, is likely more characteristic of the ISM (i.e., ${\sim}0.1{-}1$ cm$^{-3}$). However, the CSM in the nearer vicinity of the WD could be denser, as suggested by more recent type Ia observations that display possible evidence of CSM interaction (e.g., \citealt{hamuy03, mccully22}), and because the ejecta from the weak supernova constitutes an additional reservoir of material.} of ${\sim}10^{2}$ cm$^{-3}$, and hence a mass density of $\rho_{\rm a} \simeq 10^{-22}$ g cm$^{-3}$. Note that the evolution is self-similar when the density contrast between the wind and the ambient medium is high. Therefore, choosing a larger value of $R_{\rm i}$ while keeping the same normalization of the density results in the same solution but at a later time. Alternatively, adopting a larger value of the ambient density while keeping the same value of $R_{\rm i}$ results in the same solution but at an earlier time. At $R_{\rm i} = 10^{15}$ cm, the wind density given in Equation \eqref{rhow} drops to $\rho_{\rm w} \simeq 10^{-19}$ g cm$^{-3}$, corresponding to an initial wind-to-ambient density ratio of $\rho_{\rm w} / \rho_{\rm a} \sim 10^{3}$. Going forward, we use $V_{\rm w}$, $R_{\rm i}$, $\rho_{\rm a}$, $\rho_{\rm w}$ and their estimated values above as constants that represent the initial conditions at the CD.

The initial interaction between the wind and ambient medium is therefore in the regime where $\rho_{\rm wind} \gg \rho_{\rm amb}$. This interacting stage of the SNR evolution lasts until $\rho_{\rm wind} \simeq \rho_{\rm amb}$, corresponding to when the wind expands by a factor of ${\sim}10{-}100$ (i.e., to $R \sim 10^{16{-}17}$ cm). With the following wind and ambient density profiles,
\begin{equation}
    \rho_{\rm wind}(R) = \rho_{\rm w}\left(\frac{R}{R_{\rm i}}\right)^{-2}, \quad \rho_{\rm amb}(R) = \rho_{\rm a},
\end{equation}
the radial location of the CD approximately satisfies
\begin{equation}
    R_{\rm c}(t) = R_{\rm w}(t)\left(1-0.395\sqrt{\frac{\rho_{\rm a}}{\rho_{\rm w}}}\frac{R_{\rm c}(t)}{R_{\rm i}}\right), \label{Rcoft}
\end{equation}
where $R_{\rm w}(t) = R_{\rm i}\left(1+V_{\rm w}t/R_{\rm i}\right)$, while the radii of the RS and FS are, respectively,
\begin{equation}
    \begin{split}
        &R_{\rm r}(t) = R_{\rm c}(t)\left(1-0.0484\sqrt{\frac{\rho_{\rm a}}{\rho_{\rm w}}}\frac{R_{\rm c}(t)}{R_{\rm i}}\right), \\
        &R_{\rm s}(t)= R_{\rm c}(t)\left(1.0957+0.0334\sqrt{\frac{\rho_{\rm a}}{\rho_{\rm w}}}\frac{R_{\rm c}(t)}{R_{\rm i}}\right). \label{Rsoft}
    \end{split}
\end{equation}
The numerical factors come from Table 1 in \citet{coughlin24}. Note that Equation \eqref{Rcoft} is an implicit relationship for $R_{\rm c}(t)$, where the second term in parentheses is the first in an infinite series (in this case the next term is $\propto R_{\rm c}^2$), and $R_{\rm r}(t)$ and $R_{\rm s}(t)$ are written relative to $R_{\rm c}(t)$.

We assume the ambient medium is homogeneous and confined to ${\sim}$sub-pc scales (i.e., interior to the ISM), such that its density does not decline much with distance from the central WD driving the wind.\footnote{Section \ref{sec:discussion} discusses the importance of the spatial profile of the ambient medium.} In the limit of infinite density contrast, any small-scale inhomogeneities/clumps in the ambient medium will excite perturbations that  evolve largely independently of perturbations to the CD/shocked wind. In this limit, an analysis of the perturbations imparted to the shocked ambient gas is more tractable (than is the case for the shocked wind) because they are separable in $\log(t)$ and the spatial self-similar variable (e.g., \citealt{ryu87, goodman90}); it is straightforward to show that they are stable with the eigenmodes decaying with time approximately as $\propto t^{-3}$. Therefore, inhomogeneities in the shocked ambient medium will largely produce damped acoustic oscillations (cf.~\citealt{coughlin22}) that do not significantly modify the shocked wind, although they could still seed the RTI if they are large.

\subsection{Stability and growth}
\label{sec:stability}
\noindent
The standard RTI dispersion relation is derived by considering hydrostatic upper (here ``upstream'') and lower (``downstream'') fluids, which have respective densities $\rho_{\rm u}$ and $\rho_{\rm d}$, separated by a CD. The system is in equilibrium in the presence of a uniform gravitational field $g$, such that the unperturbed upstream and downstream pressure profiles, $p_{\rm u}$ and $p_{\rm d}$, depend on displacement from the CD. Writing the perturbations as $\propto e^{i\sigma t+ik x}$, where $x$ is the position along the unperturbed CD, then leads to the following dispersion relation:
\begin{equation}
\sigma_{\rm RT}^2 = \frac{\frac{\partial p_{\rm u}}{\partial r}-\frac{\partial p_{\rm d}}{\partial r}}{\rho_{\rm u}+\rho_{\rm d}}k. \label{dispgen}
\end{equation}
Here $r$ is the displacement away from the CD, defined such that into the upstream fluid is positive, and the quantities on the right-hand side are evaluated at the location of the unperturbed CD (i.e., $r = 0$). Substituting $\partial p_{\rm u}/\partial r = -g\rho_{\rm u}$ and $\partial p_{\rm d}/\partial r = -g\rho_{\rm d}$ then yields the standard result, which in the limit of an infinitely large density contrast is $\sigma_{\rm RT}^2 = -g k $.

Although a wind impacting an ambient medium is distinct from the setup that leads to Equation \eqref{dispgen}, we can still use it to obtain a crude sense of how the RTI growth rate depends on the system properties. When the density contrast is large, such that $\rho_{\rm d} \ll \rho_{\rm u}$ and $\partial p_{\rm d}/\partial r \ll \partial p_{\rm u}/\partial r$, we can use the self-similar solutions in \citet{coughlin24} to evaluate $\rho_{\rm u}^{-1}\partial p_{\rm u}/\partial r$ near the CD. For a wind impacting a constant-density medium, this yields (from Appendix B of \citealt{coughlin24})
\begin{equation}
\sigma_{\rm RT} \simeq 
\frac{V_{\rm c}}{\sqrt{R_{\rm i}}}\left(\frac{\rho_{\rm a}}{\rho_{\rm w}}\right)^{1/4}k^{1/2}, \label{dispwind}
\end{equation}
where $V_{\rm c} \simeq V_{\rm w}$ is approximately constant when the density contrast is high. If we further let $k \rightarrow k\times R_{\rm i}$, such that $k$ is measured in units of $R_{\rm i}^{-1}$, then the growth timescale is
\begin{equation}
    \tau_{\rm RT} \simeq \frac{1}{\sigma_{\rm RT}} \simeq \frac{R_{\rm i}}{V_{\rm w}}\left(\frac{\rho_{\rm w}}{\rho_{\rm a}}\right)^{1/4}k^{-1/2} \label{tgrow}.
\end{equation}
Note that unlike the standard RTI, the dispersion relation above depends on the density contrast. This arises because the pressure gradient that drives the RTI is established by the CD deceleration, which itself depends on the density ratio.\footnote{We discuss this feature and its implications in Section \ref{sec:conclusions}.} Although the density contrast acts to slow the RTI growth, $\tau_{\rm RT} \propto \left(\rho_{\rm w}/\rho_{\rm a}\right)^{1/4}$, it more effectively slows the KHI growth, $\tau_{\rm KH} \propto \left(\rho_{\rm w}/\rho_{\rm a}\right)^{1/2}$.

\subsection{Filament properties}
\label{sec:properties}
\noindent
The RTI should stop operating roughly when the wind and ambient densities are equal;\footnote{Strictly speaking it is also when there is no longer a strong pressure gradient in the shocked wind relative to the shocked ambient medium, which should occur roughly simultaneously with when the densities become comparable, as at this point the FS decelerates more substantially and the solution enters the self-similar phase described in \citet{weaver77}.} that is, when $R_{\rm c}/R_{\rm i} = 10{-}100$, or $R_{\rm c} = 10^{16-17}$ cm (see Section \ref{sec:parameters}). Reaching this equality takes just ${\sim}$1--10 yr for a wind speed of $10^{3{-}4}$ km s$^{-1}$. The RTI growth timescale of the filaments is shortest for the smallest-scale perturbations (i.e., largest $k$; see Section \ref{sec:stability}). Consequently, the filaments should be highly elongated with a length that is limited by the timescale over which the RTI operates.

We expect the RTI plumes to accelerate downstream until the ram pressure of the medium into which they advance (i.e., as measured in the frame of the CD) balances their thermal pressure (this is equivalent to the condition that their speeds be subsonic with respect to the downstream material), which---if their speed is subsonic in the frame of the CD---should be comparable to the pressure at the CD. The firm upper limit to the speed of the RTI plumes relative to the CD as they propagate downstream into the shocked ambient medium is then
\begin{equation}
    v_{\infty, \rm d} \simeq \sqrt{\frac{p_{\rm CD}}{\rho_{\rm d}}} \simeq 0.46 V_{\rm c}, \label{vinf}
\end{equation}
where the numerical factor of $0.46$ arises from the assumption that the FS shell is gas-pressure dominated (it is $0.36$ for a radiation-dominated gas). This is faster than the initial speed of the FS, which is ${\sim}0.1 V_{\rm c}$ (again relative to the CD; see Equation \ref{Rsoft}), implying that the RTI plumes could overtake the FS (as seen in simulations; \citealt{duffell24, blondin01}). That said, Equation \eqref{vinf} is likely a substantial overestimate because it crudely assumes uniform pressure within the plumes, despite their implied velocity being supersonic.\footnote{Due to the large density contrast, the shocked wind sound speed is $\sqrt{\rho_{\rm a}/\rho_{\rm w}} V_{c} \ll 0.46 V_{c}$; thus, the plumes are supersonic.} Therefore, the filaments likely propagate downstream only marginally faster than the CD, whose lab-frame speed $V_{\rm c}$ is of order the natal WD wind speed (${\sim}{\rm few} \times 10^{3}$ km s$^{-1}$; see Section \ref{sec:parameters}) and is consistent with the observed speed of the filaments \citep{cunningham24}.

Applying the same reasoning, the asymptotic upstream speed of the RTI bubbles is (relative to the CD)
\begin{equation}
    v_{\infty, \rm u} \simeq \sqrt{\frac{p_{\rm CD}}{\rho_{\rm u}}} \simeq \sqrt{\frac{\rho_{\rm a}}{\rho_{\rm u}}} v_{\infty, \rm d}. 
    \label{vu}
\end{equation}
This speed is subsonic with respect to the shocked ambient gas; thus, it is more self-consistent than Equation \eqref{vinf} and much slower than $v_{\infty, \rm d}$. The RTI plumes therefore travel downstream into the shocked ambient medium faster than the RTI bubbles travel upstream into the shocked wind, as seen in Figure \ref{fig:rti}. Equation \eqref{vu} is comparable to the initial speed of the RS relative to the CD speed, where the relative difference between the two is $\simeq 0.1\sqrt{\rho_{\rm a}/\rho_{\rm w}}$, although the RS speed grows as $\propto R_{\rm c}$ \citep{coughlin24}. Therefore, the RTI bubbles may overtake and disrupt the RS, depending on how quickly the RTI grows and the magnitude of the initial perturbations.

The wind-fed filaments advance into the shocked ambient gas marginally faster than the CD. Therefore, the velocity shear $\Delta{v}$ is small and remains so even after the RTI halts, whereas the density contrast remains large because the filaments expand uni-directionally. The KHI growth timescale is then much longer than the age of Pa 30, $\tau_{\rm KH} \sim \sqrt{\rho_{\rm wind} / \rho_{\rm amb}} / (k \Delta{v}) \gg 850~{\rm yr}$, as follows from the following conservative estimates \citep[e.g.,][]{cunningham24}: a current density contrast of ${\sim}100$; a wavenumber corresponding to the current filament width of ${\sim}0.01$ pc; and, because the filaments expand with the FS shell, a $\Delta{v}$ much slower than the current filament speed of ${\sim}10^{3}~{\rm km~s^{-1}}$. Therefore, as observed, we expect the filaments to remain coherent out to the current epoch.

From the considerations above, the filaments should expand primarily into the shocked ambient medium; grow predominantly in the radial direction, because their large densities and low shear inhibit efficient KHI growth; and decline in density as ${\sim}1/r$, because they are pressure confined. Therefore, we expect the RTI-induced filaments to have large aspect ratios and to extend approximately radially outward from the system center. If the wind density was ${\sim}10^{-19}$ g cm$^{-3}$ at $R_{\rm i} = 10^{15}$ cm (see Section \ref{sec:parameters}), then after radially expanding by a factor of ${\sim}100$ (at which point the RTI should cease), their densities would fall to ${\sim}10^{-21}$ g cm$^{-3}$. This corresponds to a number density of ${\sim}{\rm few} \times 100$ cm$^{-3}$, which is consistent with those observed of ${\sim}200{-}1000$ cm$^{-3}$ based on [S II] emission lines \citep{fesen23}.

The initial temperature of the shocked (assumed gas-pressure-dominated) wind is $T \simeq 10^{5{-}6}$ K, which arises from a density contrast of $\rho_{\rm w}/\rho_{\rm a} \simeq 10^{3}$ and a wind speed of ${\sim}{\rm few} \times 10^{3}$ km s$^{-1}$ (see Section \ref{sec:parameters}). If the filaments grow adiabatically, then their temperature scales as $\rho^{2/3}$ and falls to ${\sim}10^{3{-}4}$ K by the time they expand by a factor of ${\sim}10^{2{-}3}$ to reach pc-scale lengths (in line with inferences in \citealt{cunningham24}). Therefore, we expect the filaments to be optically bright but largely devoid of hydrogen (because they originate from a WD-driven wind at early epochs), which is consistent with observations.

\section{Discussion}
\label{sec:discussion}
Pa 30 is seemingly the only known SNR with a firework-like morphology, which begs the question of the origin of its uniqueness. As noted in the introduction, its high-speed WD wind offers a suggestive clue. Our model incorporates an early-phase WD wind, driven by either radioactive decay of freshly synthesized elements \citep{shen17} or supercritical disc accretion (see Section \ref{sec:conclusions}). We propose that the RTI generates tendrils of high-density, shocked wind material that propagate into the low-density, shocked ambient medium. The high mass-loss rate from the natal WD wind implies that its density remains large relative to the CSM out to large distances. Crucial to our proposed RTI filament formation mechanism, the system thus maintains a high density contrast without quickly transitioning to the self-similar solution described in \citet{weaver77}; as the nearly isobaric shocked wind would be RTI-stable in this wind-driven analog of the Sedov-Taylor blast wave. Presumably, other supernovae also go through a phase of high density contrast (e.g., \citealt{bjornsson25}), yet they do not display a firework morphology. In some cases, this phase is likely too short-lived for filaments to develop. For instance, type Ibc and type II supernovae enshrouded in a substantial gaseous reservoir (e.g., from late-stage stellar winds and eruptions) likely lack the necessary large density contrast extending out to large distances.

The morphology could also relate to the pressure gradient in the shocked wind arising from the CD deceleration, which depends on the time-dependent density contrast---this differs from the standard RTI, where a time-independent gravitational field generates the pressure gradient. This effect manifests in the approximate dispersion relation given by Equation \eqref{dispwind}, where the density ratio enters to inhibit the RTI growth rate, as $\propto (\rho_{a}/\rho_{w})^{1/4}$; albeit less effectively than it inhibits the KHI growth rate, as $\propto (\rho_{a}/\rho_{w})^{1/2}$.
 
Considering more general setups, such that either a wind ($m = 2$) or homologous ejecta ($m = 3$) drives the explosion and the ambient medium has a power-law profile with index $n$ (i.e., $\rho_{a} \propto r^{-n}$), we find Equation \eqref{dispwind} generalizes\footnote{We caution against over-interpreting this generalization because the additional time dependence implies the solutions are not simply exponentials in time.} to $\sigma_{RT} \rightarrow \sigma_{RT} \times t^{(m-n)/4-1/2}$, where we approximated the CD position as $R_{\rm c} \propto t$. Scenarios where a wind encounters a medium with declining density (e.g., a wind from a protostar running into a Bondi-like medium with $n = 3/2$) should thus be less susceptible to the RTI than the constant-density case we considered. If $n = 2$, then there is no deceleration, and if $n > 2$ the pressure profile in the shocked wind inverts (see the bottom-left panel of Figure 2 in \citealt{coughlin24}); in these cases the RTI would not be expected to operate, so no tendrils form. Compared to a wind-driven explosion, this generalization suggests that homologously expanding explosions are more susceptible to RTI, due to the enhanced deceleration rate of the CD (and hence stronger pressure gradient). However, the more rapidly declining ejecta density profile in this scenario serves to shorten the timescale over which the RTI operates and renders the tendrils more susceptible to subsequent KHI; the enhanced deceleration of the CD further reduces the growth rate of the instability (Equation \ref{dispwind} depends on $V_{\rm c}$). Plausibly, the early phases of a homologous explosion (i.e., prior to the density ratio satisfying the condition of \citealt{chevalier82}) are too short-lived for RTI-induced spikes to lengthen into tendril-like structures before KHI sets in.

When the RTI channels a substantial fraction of the shocked wind into long, thin tendrils, it dramatically reduces the wind's average effective density. This reduction in density, coupled to the simultaneous upstream-moving RTI-induced plumes of shocked ambient gas, inhibits the outward propagation of the RS. When the driving mechanism switches, presumably to magnetic \citep[e.g.,][]{gvaramadze19, kashiyama19, zhong24}, the wind speed increases and the mass-loss rate declines, such that the wind density drops below that of the ambient gas and the RS speed slows substantially compared to the wind speed. The subsequent evolution of the wind termination shock could then plausibly be described by the \citet{weaver77} self-similar phase, during which $R_{\rm r} \propto t^{2/5}$. If we assume the current wind speed and mass-loss rates from \citet{oskinova20} were valid over the entire age of the remnant (but after the initial phase that formed the filaments), then the current position of the RS after 850 years---using Equation (10) in \citet{weaver77}---would be ${\sim}0.07$ pc, which is a factor of $\sim$10 smaller than the observed 0.6 pc inner edge of the filament-less cavity \citep{fesen23, cunningham24}. In this model, it therefore seems more likely that the inner cavity traces the time-dependent location of the trailing edges of the filaments when the RTI stopped operating, as opposed to the wind termination shock.

The Pa 30 filaments display faint [O III] emission \citep{cunningham24} but the WD and its wind are currently oxygen-rich \citep{oskinova20}. If the filaments are made of WD wind material, then why are these oxygen levels so different? As noted above, during the filament formation epoch, the wind-driving mechanism was probably related to the radioactive decay of newly synthesized elements following the merger-induced deflagration, which failed to produce a detonation \citep{shen17}, or to a hypercritical phase of accretion (Section \ref{sec:conclusions}). Therefore, during the short time that the filaments formed and elongated ($\sim$1--10 years; Section \ref{sec:properties}), the natal wind was likely polluted by heavy elements; thus, explaining their low oxygen content.

Our simple hydrodynamic simulations (see Figure \ref{fig:rti}) used a single-mode perturbation to seed the RTI, leading to spikes of equal length and spacing. A more stochastic distribution of fluctuations within the ejecta would produce a more varied distribution of tendril lengths and widths. Ambient medium fluctuations, if sufficiently strong (see end of Section \ref{sec:parameters}), could also contribute to tendril variety. Given that the RTI growth rate likely scales with spherical harmonic as $\sigma_{\rm RT} \propto \ell^{1/2}$ at sufficiently high $\ell$ (i.e., as the standard RTI analysis predicts; see Equation \ref{dispgen}), one might expect narrower filaments to form first but then truncate (due to the onset of wider filaments sapping the mass supply), leading to a small-scale clumpy structure. However, this outcome depends on complex couplings involving the non-linear instability growth, the evolving RS shell width, and extensions to the standard RTI analysis. Imminent imaging of Pa 30 by the \textit{James Webb Space Telescope} and the \textit{Hubble Space Telescope} will provide further insight into the coherence of the filaments.

\section{Summary and Conclusions}
\label{sec:conclusions}
\noindent
We proposed an explanation for the firework-like morphology of the Pa 30 SNR. The natal WD drives a high-density wind into the CSM, whose density is low and declines gradually compared to a homologous explosion (Section \ref{sec:parameters}). The large density contrast at the CD enables the RTI to grow with an approximate $e$-folding timescale of $\tau_{\rm RT} \sim \left(R_{\rm i}/V_{\rm w}\right)\times\left(\rho_{\rm a}/\rho_{\rm w}\right)^{1/4}\times k^{-1/2}$, with $k$ in units of $R_{\rm i}^{-1}$ (Section \ref{sec:stability}). A continuous supply of shocked wind material gets channeled into the RTI spikes as they elongate toward the FS, all the while draining the reservoir of shocked wind material in the RS shell.\footnote{The compounding effect of low-density bubbles (made of shocked ambient gas) moving into the RS shell further lowers its density.} Consequently, the CD reaches density parity after just ${\sim}$1--10 yr and the RTI ceases, which truncates the spikes (Section \ref{sec:properties}). The spikes remain intact because their large density contrast with the shocked CSM and low shear velocity---a consequence of the FS moving at nearly the same speed as the CD---prevent their destruction by KHI (Section \ref{sec:properties}).

At a high level, our model suggests geometric, kinematic, and thermodynamic properties of the filaments that are consistent with observations (Sections \ref{sec:properties} and \ref{sec:discussion}). An important aspect of our model is that the dense wind of the natal WD remnant---not the preceding type Iax supernova ejecta---primarily dictates the firework morphology of Pa 30, in line with radio upper limits that imply a weak explosion \citep{shao25}. Similar morphologies could be present in other type Iax systems, especially if a prominent feature of their evolution is a dense wind encountering a rarefied CSM (as suggested by, e.g., \citealt{camacho23, callan24, kwok25, schwab25}).

Supernovae should be susceptible to the RTI and KHI during various phases of their evolution \citep{chevalier76a}. These processes and their development have been studied with hydrodynamical simulations (e.g., \citealt{chevalier78, fryxell91, dwarkadas00, stone07, miles09, garcia12, ferrand22, prust25}), often assuming the self-similar solutions described in \citet{chevalier82} act as the unperturbed background state (e.g., \citealt{chevalier92, jun96, blondin01, fraschetti10, duffell17, mandal23}). These solutions are known to be unstable \citep{chevalier92}, require time-independent ejecta-to-ambient density ratios of order unity (see Table 1 in \citealt{chevalier82}), and are thus prone to both the RTI and subsequent KHI (see, e.g., Figure 1 in \citealt{duffell17}). However, the scenario we described that leads to the RTI (sans KHI) is qualitatively distinct from the \citet{chevalier82} phase and is less straightforward to numerically simulate because the relative separation between the CD and the RS is $\propto \sqrt{\rho_{\rm a}/\rho_{\rm ej}}$ \citep{hamilton84, truelove99}. Thus, a large density contrast confines the shocked wind to a thin shell, as does the small coefficient of proportionality (${\sim}{\rm few} \times 0.01$), which in our case of a wind impinging on a constant-density medium is $0.0484$ \citep{coughlin24}.

\begin{figure*}
    \includegraphics[width=0.455\textwidth]{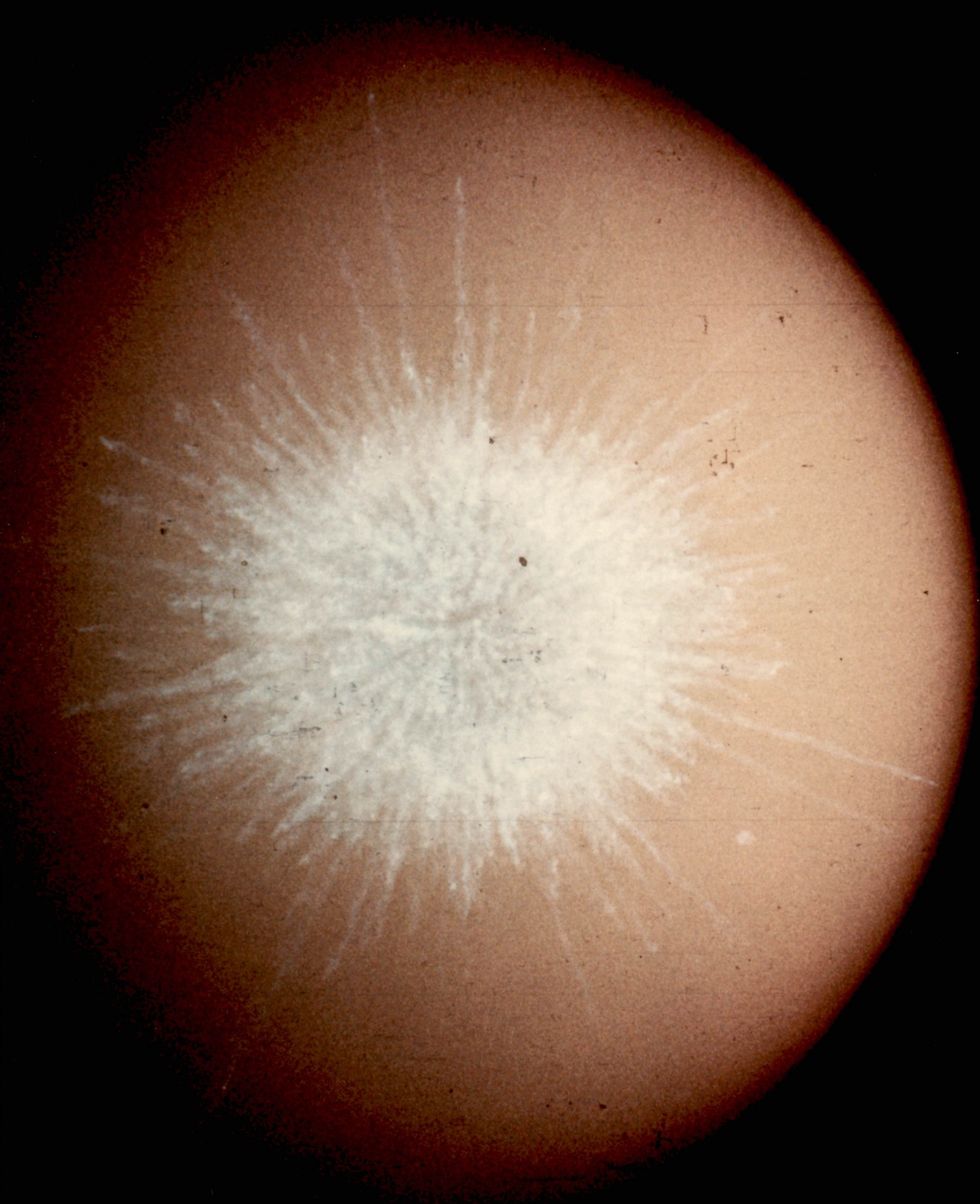}
       \includegraphics[width=0.525\textwidth]{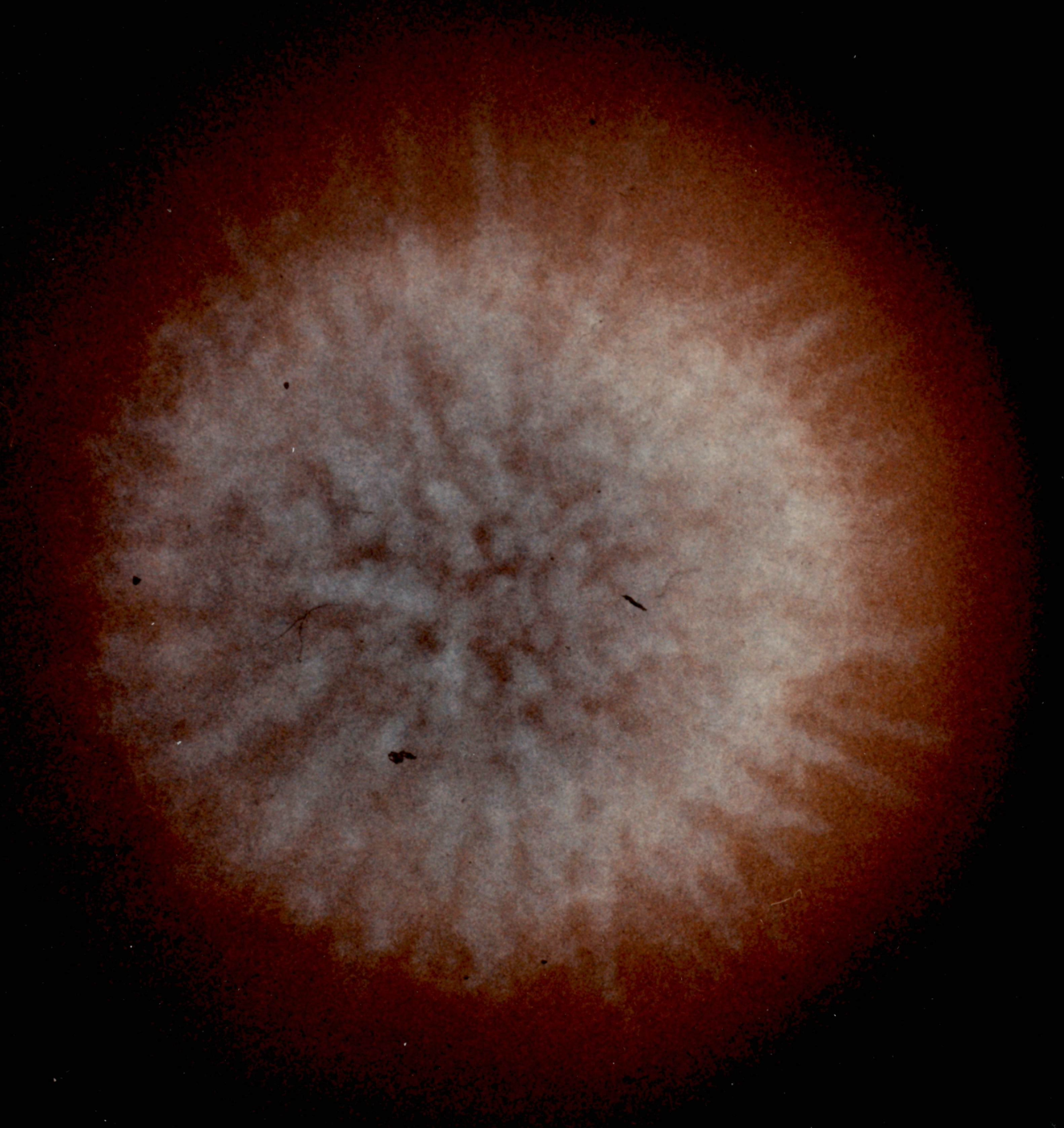}
     \caption{Photographs of the ``Kingfish'' high-altitude nuclear test (LA-UR-25-31831). The left image was taken ${\sim}40$ ms post-detonation, illustrating the formation of tendril-like filaments in the interior of the explosion. The right image was taken ${\sim}256$ ms post-detonation, showing the transformation of the firework-like filaments into a more cauliflower-like morphology reminiscent of standard supernova remnants. This terrestrial experiment suggests that homologous explosions may also pass through a Pa 30-like evolutionary phase that is (compared to wind-driven explosions) short-lived.}
     \label{fig:explosion}
\end{figure*}

We also expect the dispersion relation to change qualitatively once the perturbation length scale is larger than the separation between the RS and CD.\footnote{If one accounts for the finite vertical extent of the upper fluid, $H$, the standard RTI dispersion relation becomes $\sigma_{\rm RTI}^2 = \frac{\rho_{\rm d}-\rho_{\rm u}}{\rho_{\rm u}\tanh(|k|H)+\rho_{\rm d}}|k|g$, implying that the growth rate is suppressed once $|k|H\lesssim \rho_{\rm d}/\rho_{\rm u}$; since $H \propto \sqrt{\rho_{\rm d}/\rho_{\rm a}}$, this translates to $k \lesssim 1$ when $k$ is measured in units of $R_{\rm i}^{-1}$.} If the initial encounter between the wind and the ambient medium takes place at $R_{\rm i} = 10^{15}$ cm with a density contrast of $10^{3}$, then the initial width of the shocked wind layer would be ${\sim}0.0484 \times 10^{-3/2} R_{\rm i} \simeq 10^{12}$ cm. If the maximum transverse size of RTI-unstable modes were limited to this length scale (to within a factor of ${\sim}\pi$), the filament aspect ratio would decline from $1/1000$ with time, implying they would be unobservable (not to mention much narrower than their observed aspect ratio of ${\sim}0.01{-}0.1$; \citealt{cunningham24}). However, it is unclear if $R_{\rm i}$ alone sets the width of the filaments, or if their widths expand as the CD radius and RS shell width expand. Unfortunately, the methodology in Section \ref{sec:stability} cannot differentiate between these possibilities---the WKB approximation implicitly assumes that the growth timescale is much shorter than $R_{\rm i}/V_{\rm w}$. We intend to explore the nature of the RTI during this phase both numerically and analytically in future work.

As previously noted, radio observations of Pa 30 imply that the 1181 AD type Iax supernova was of exceptionally low energy \citep{shao25}, and the spectroscopically inferred elemental abundances of the central star suggest that the progenitor consisted of the merger of a high-mass ONe WD (mass $\gtrsim 1~M_{\odot}$) and a lower-mass CO WD \citep{oskinova20}. The merger therefore likely proceeded through the tidal disruption of the CO WD (e.g., \citealt{kashyap18}), resulting in the formation of an accretion flow around the ONe WD disruptor and leaving much of the explosion ejecta bound to the system to fuel additional, later-time fallback. Therefore, instead of the initial wind being launched by the delayed radioactive decay of heavy elements \citep{shen17}, it may have been driven by a period of sustained supercritical accretion with mass loss rates presumably much larger than assumed in Section \ref{sec:parameters}. Similar physical conditions (an overdense wind impacting a medium) may be present in, e.g., the collapsar model of long gamma-ray bursts \citep{macfadyen99}, delayed circularization and black-hole disc formation in failed supernovae \citep{quataert19}, hyperaccreting neutron stars/ultra-luminous X-ray sources (e.g., \citealt{strohmayer03}), and the winds from tidal disruption events (e.g., \citealt{miller15}), to which it would be interesting to apply our model---both in the context of the wind-CSM dynamics and the possibility of filament formation.

{Finally, while Pa 30-like supernova remnants are seemingly rare, terrestrial explosions provide another route for testing the hypothesis that tendril-like structures can form during the early phases of explosions (as suggested by the generalized dispersion relation in Section \ref{sec:discussion}). Figure \ref{fig:explosion} gives one such example from the ``Kingfish'' high-altitude nuclear test conducted by the United States in 1962: the left panel, which is a photograph of the explosion taken ${\sim}40$ ms following the initial detonation, illustrates clear evidence of radial filaments that developed. The right panel shows a photograph of the same explosion, but ${\sim}256$ ms post-detonation, showing the filaments starting to evolve into more of a cauliflower-like structure. This result suggests that homologous explosions (i.e., more typical supernovae) may pass through a morphological phase more similar to that of Pa 30, but with a temporal duration that is more short-lived than that of a wind-driven explosion.}

\section*{}
We thank the anonymous referee for a very useful report that added to the quality and clarity of this work. We also thank Eliot Quataert for constructive comments, and Chris Nixon for providing feedback on an early version of the manuscript. E.R.C.~acknowledges support from the National Aeronautics and Space Administration through the Astrophysics Theory Program, grant 80NSSC24K0897. Los Alamos National Laboratory approved this paper for unlimited release (LA-UR-25-31324).

\bibliographystyle{aasjournalv7}

\end{document}